\documentclass[10pt,conference]{IEEEtran}
\usepackage{epsfig,setspace,amsmath,epsf,amssymb,bm,theorem,cite,graphicx, epstopdf,algorithm,algpseudocode,float,color,mathtools,authblk,tikz,physics}
\usepackage{enumitem} 
\usepackage{optidef}
\usepackage{booktabs}
\usepackage{subfigure}
\usepackage{bbm}
\usetikzlibrary{positioning}
\usetikzlibrary{decorations.text}
\usetikzlibrary{decorations.pathmorphing}

\newcommand{\mycomment}[1]{}

\newtheorem{theorem}{Theorem}

\newenvironment{Proof}[1]{\medskip\par\noindent{\bf Proof:\,}\,#1}{{\mbox{\,$\blacksquare$}\par}}

\IEEEoverridecommandlockouts
\allowdisplaybreaks
\begin{document}

\title{Optimal Source Coding of Markov Chains for \\ Real-Time Remote Estimation}
\author{Ismail Cosandal \qquad Sennur Ulukus\\
\normalsize Department of Electrical and Computer Engineering\\
\normalsize University of Maryland, College Park, MD 20742\\
\normalsize  \emph{ismailc@umd.edu} \qquad \emph{ulukus@umd.edu}}

\maketitle

\begin{abstract}
    We revisit the source coding problem for a Markov chain under the assumption that the transmission times and how fast the Markov chain transitions its state happen at the same time-scale. Specifically, we assume that the transmission of each bit takes a single time slot, and the Markov chain updates its state in the same time slot. Thus, the length of the codeword assigned to a symbol determines the number of non-transmitted symbols, as well as, the probability of the realization of the next symbol to be transmitted. We aim to minimize the average transmission duration over an infinite horizon by proposing an optimal source coding policy based on the last transmitted symbol and its transmission duration. To find the optimal policy, we formulate the problem with a Markov decision process (MDP) by augmenting the symbols alongside the transmission duration of the symbols. Finally, we analyze two Huffman-based benchmark policies and compare their performances with the proposed optimal policy. We observe that, in randomly generated processes, our proposed optimal policy decreases the average transmission duration compared to benchmark policies. The performance gain varies based on the parameters of the Markov process.
\end{abstract}

\section{Introduction}
The source coding problem for a Markov source dates back decades \cite{ott2003compact}. The most conventional way is to encode multiple symbols as a block, and to apply a trellis-like algorithm to decode \cite{ott2003compact, bahl2003optimal}. This coding strategy introduces an inherent delay due to the block length; thus, it is not suitable for real-time monitoring problems. An alternative approach, namely zero-delay transmission, considers encoding each symbol of the process separately without waiting for the whole block \cite{witsenhausen1979structure, phamdo2002optimal}. The trade-off between a shorter block size and a higher bit error rate is investigated in \cite{phamdo2002optimal} for binary symmetric channels.

In more recent works \cite{cregg2024reinforcement, cregg2024near}, a near-optimum source coding method is derived with the aid of reinforcement learning. In these works, each symbol of a Markov source is encoded with a quantization function whose output alphabet is smaller than the source alphabet, resulting in a distortion. Since the remote monitor is aware of the quantization function, it maintains the probability of distribution of the source process, which is referred to as the \emph{belief}. These papers aim to associate each belief with a quantization function; however, beliefs are continuous-valued, hence the state space of this problem is not finite. In \cite{cregg2024reinforcement}, the belief is quantized, and the near-optimum policy is obtained over the quantized beliefs. In \cite{cregg2024near}, on the other hand, a finite memory is utilized to discretize the state space. The zero-delay framework ignores the additional delay caused by the transmission time, which is an asymptotic approach as addressed in \cite{cregg2024near}. 

In this paper, we investigate the case where the transmission duration is proportional to the codeword length, and the Markov process continues to evolve during the transmission. In other words, we consider a case where the transmission of each bit takes a single slot; thus, some updates cannot be transmitted for codeword lengths larger than $1$. As a result of this, the probability of the next encoded symbol depends on the codeword length of the previous symbol. This assumption is practical for applications where the source process evolves at the same time-scale as the channel delay, due to either fast-changing source processes or slow-rate channels.

In recent works \cite{optimal_codes, bastopcu2021selective, chen2019benefits, singhvi2023coding, Wei_chen, cosandal2025timely, liyanaarachchi2025source}, the effect of codeword length on the performance of real-time monitoring problems has been investigated for various settings. In \cite{optimal_codes}, a source generates independent and identically distributed (i.i.d.) symbols in each time slot, and each symbol is encoded with a predetermined codebook. Similar to our assumption in the current paper, in \cite{optimal_codes}, choosing a longer codeword length causes a delay in transmission, resulting in the transmitted update becoming stale and also not transmitting the updates generated by the source during the ongoing transmission. \cite{optimal_codes} derives an optimum codebook that minimizes the freshness metric of age of information (AoI) \cite{Yates__HowOftenShouldone}, which is widely used for real-time monitoring problems \cite{yates2020age}. Similarly, in \cite{bastopcu2021selective}, it is shown that a lower AoI can be achieved by not transmitting symbols with low probabilities. In addition, \cite{chen2019benefits} and \cite{singhvi2023coding} propose AoI-minimizing source coding for multi-source system models.

The works \cite{Wei_chen, cosandal2025timely, liyanaarachchi2025source} study the remote source coding of a Wiener process to minimize the mean-squared error (MSE). In \cite{Wei_chen}, the increment of the Wiener process is encoded with a high-rate quantization. To address accumulated quantization error on the receiver side, \cite{Wei_chen} proposes a multi-level error correction scheme that periodically quantizes the accumulated error from the previous level alongside the increment of the process. In \cite{cosandal2025timely}, the Wiener process is encoded periodically with a very low-bit (single-bit) quantization strategy. In addition to the incremental process, the previous quantization error is also encoded using a dynamic quantization function derived from the history of previously transmitted symbols, thereby preventing the accumulation of error. Another recent study \cite{liyanaarachchi2025source} considers a lossless transmission scheme with four dynamic thresholds and the generate-at-will approach. For each threshold, a real-valued codeword length is assigned. The increment of the process is sampled whenever it reaches one of the thresholds, and the monitor receives the sampled process after a deterministic delay based on the length of the corresponding threshold's codeword.

The paper \cite{bobbili2020variable} proposes encoding symbols using two different codeword lengths for a Markov process, and analyzes AoI-performance of this strategy. For each symbol, the sensor transmits either the actual update or encodes the difference from the previous transmitted symbol with an incremental update by using a shorter codeword based on the amount of the increment. In \cite{poojary2019real}, both true and incremental updates are encoded by the same-size codewords by inserting more parity bits for incremental updates. This strategy allows transmitting small increments with lower probability of error.

The contribution of our paper can be summarized as follows: 
\begin{itemize}
    \item We study the source coding problem for a Markov chain by assuming that the transmission of each bit takes a single time slot, and the Markov process updates its state during the transmissions. To the best of our knowledge, this is the first paper studying the optimal source-coding problem under this assumption.
    \item We formulate the problem as a Markov decision process (MDP) by augmenting the state space with the current transmission durations, and then find the optimal policy using the policy iteration algorithm.
    \item We additionally investigate two Huffman coding-based benchmark policies. We derive analytical results for these policies and compare their performance with the performance of the proposed optimal policy.
\end{itemize}

\begin{figure}
    \centering
    \includegraphics[width=0.95\linewidth]{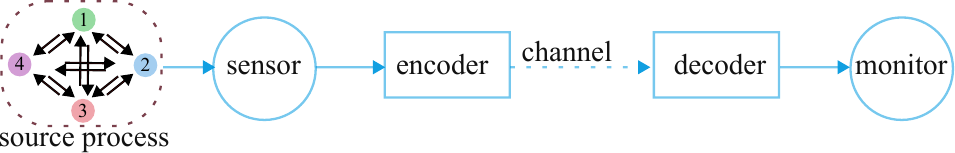}
    \caption{Illustration of the system model for an example process with $N=4$.}
    \label{fig:sys}
\end{figure}

\emph{Notation:} We use the following notation throughout the paper. ${a}_{m}$ denotes the $m$th entry of the vector $\bm{a}$, and $\bm{A}_{mn}$ or $\left(\bm{A}\right)_{mn}$ denotes the $(m,n)$th element of the matrix $\bm{A}$. We denote the $n$th row of matrix $\bm{A}$ by $\big(\bm{A}\big)_{n*}$. The matrix exponential function is defined as $\mathrm{Exp}({\bm{A}})=\sum_{k=0}^\infty \frac{1}{k!}\bm{A}^k$ \cite{wilcox1967exponential}. Finally, $\bm{I}$ and $\bm{1}$ are the identity matrix and a column vector of all ones, respectively.

\section{System Model and Problem Formulation}
We consider a slotted-time system, where a sensor observes a stochastic process $X_t$ which is an irreducible and aperiodic, hence ergodic, discrete-time Markov chain (DTMC) with a finite alphabet-size $N$, and the symbol space $\mathcal{N}=\{1,2,\dots,N\}$. The sensor encodes the observed symbol and transmits it to a remote monitor via an error-free channel. The received signal is decoded at the monitor. The system model is illustrated in Fig.~\ref{fig:sys}. Our main assumptions in this work are that the transmission of each bit takes a single time slot, the source process changes in each time slot with the transition matrix $\bm{P}$, and the sensor should wait to finish its transmission before taking another sample. Therefore, if the symbol $n\in\mathcal{N}$ is encoded with a codeword length $\ell$, then $\ell-1$ updates cannot be transmitted during this transmission. In this work, we aim to minimize the average transmission duration in an infinite horizon by developing an optimal source coding policy.  

\begin{figure}
    \centering
    \includegraphics[width=0.85\linewidth]{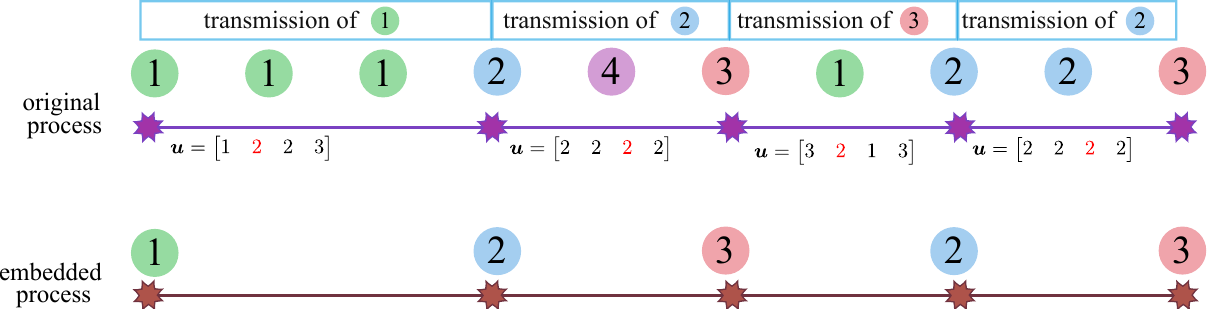}
    \caption{A sample path of a Markov chain with $N=4$ with corresponding actions, and the embedded process.}
    \label{fig:mcsm}
\end{figure}

In Fig.~\ref{fig:mcsm}, a sample path for the process $X_t$ is illustrated for $N=4$. For each symbol, both the sensor and the monitor agree on a codebook with corresponding codeword length $u_j$ for the next symbol $j$. For instance, transmission of the first symbol is encoded with a codeword length $3$, hence its transmission takes $3$ time slots. Then, the codeword lengths for the upcoming symbols, i.e., for $\{1, 2, 3, 4\}$, are chosen as $1, 2, 2, 3$, respectively. After the completion of the first symbol, the symbol $2$ is observed by the sensor, and its transmission starts with a codeword length $2$. As a result of this scheme, not all symbols can be transmitted from the original process, and transmitted symbols form an embedded process, as illustrated in Fig.~\ref{fig:mcsm}. Actions are taken only for the states of the embedded process, which we refer to as the \emph{embedded states}.

The main challenge in formulating this problem with a Markov decision process (MDP) is that each action taken in an embedded state affects the transmission time (hence the transmission probability) of the next symbol. For instance, returning to the example in Fig.~\ref{fig:mcsm}, if $u_2=1$ was selected in the first embedded state instead of $u_2=2$, the third embedded state would be $4$ instead of $3$. That violates the Markov property and disallows us from obtaining an MDP over embedded states. To overcome this problem, we propose an \emph{augmented state} approach by including transmission lengths in the state space. 

Before introducing the proposed MDP formulation, we need a definition and the following theorem. In order for a codebook to be instantaneously and uniquely decodable, codeword lengths should satisfy the Kraft's inequality \cite{cover1999elements}, 
\begin{align}
    \sum_{n\in\mathcal{N}}2^{-\ell_n}\leq 1, \label{eq:kraft}
\end{align}
where $n\in\mathcal{N}$ is the symbol encoded with a codeword length $\ell_n$. In this paper, we only consider the codebooks that satisfy the Kraft's inequality with equality, which are referred to as \emph{complete codes} \cite{mackay2003information}. It is known that in the binary-tree representation of a complete code, each node is either a leaf or has two descendant nodes \cite{cover1999elements}.

\begin{theorem} \label{thm:one} 
    The maximum codeword length of a complete code used for $N\geq2$ symbols cannot be greater than $N-1$.
\end{theorem}

\begin{figure}
    \centering
    \includegraphics[width=0.9\linewidth]{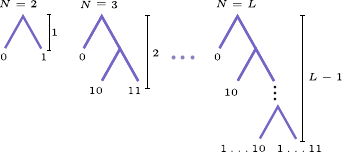}
    \caption{Binary representation of complete codes with the maximum depth.}
    \label{fig:proof}
\end{figure}

\begin{Proof}
    We use the induction method. First, notice that for $N=2$, the only complete code has codeword lengths $\ell_1=\ell_2=1$, which is consistent with the theorem. Now, assume that the theorem holds for $N$, and we have a complete code for $N$ symbols with codeword lengths $\ell_n$, $n\in\{1,\dots,N\}$, and the maximum codeword length is $N-1$. Without loss of generality, we consider that codeword lengths are ordered in ascending order for the rest of the proof, i.e., $\ell_1\leq\ell_2\dots\leq\ell_N=N-1$. From \eqref{eq:kraft} and the definition of a complete code, we have the following relation for codeword lengths,
    \begin{align}
        \sum_{n}^N2^{-\ell_n}=& 1,\\
        \sum_{n}^{N-1}2^{-\ell_n}=&1-2^{-(N-1)}. \label{eq:proof2}
    \end{align}
    Now, we consider another complete code for $N+1$ symbols. To maximize the maximum codeword length for a new complete code, we should keep $\ell_1\dots\ell_{N-1}$ the same, and only change $\ell_N$ alongside a new codeword length $\ell_{N+1}$. From \eqref{eq:proof2}, these codeword lengths should satisfy
    \begin{align}
        \sum_{n=1}^{N+1}2^{-\ell_n}=&\sum_{n=1}^{N-1}2^{-\ell_n}+\sum_{n=N}^{N+1}2^{-\ell_n}\\
        =&1-2^{-(N-1)} + \sum_{n=N}^{N+1}2^{-\ell_n}=1.
    \end{align}
    Notice that the only integer-valued codeword lengths that satisfy this condition are $\ell_N=N$, and $\ell_{N+1}=N$, which makes the maximum codeword length $N$, and completes the proof. This procedure is illustrated with Fig.~\ref{fig:proof} using a binary tree representation. 
\end{Proof}

We propose an MDP formulation with tuple $(\mathcal{S},\mathcal{U}_N,T,c)$ from \cite{ibe2013markov}, which can be summarized as:
\begin{itemize}
    \item We define the state of our problem as $s=(n,\ell)$, where $n\in\mathcal{N}$ corresponds to the symbol, and $\ell$ is the transmission length of the symbol, which is determined at the last decision point. Therefore, the state space is defined as $\mathcal{S}=\mathcal{N}\times\mathcal{L}$ where $\mathcal{L}=\{1,2,\dots,N-1\}$. Notice that we truncate the lengths by $N-1$ to represent $N$ symbols in the worst case, as stated in Theorem~\ref{thm:one}.
    \item We denote the action taken upon reaching state $s$ by $\bm{u}_s=[u_s(1),\dots,u_s(N)]$ that corresponds to the codeword length for the next symbol. We denote the action space for alphabet size $N$ by $\mathcal{U}_N$ that includes all complete codes for alphabet size $N$. For instance, the action space for $N=3$ is $\mathcal{U}_3=\left\{ \begin{bmatrix}1 & 2 & 2\end{bmatrix},\ \begin{bmatrix}2 & 1 & 2\end{bmatrix},\ \begin{bmatrix}2 & 2 & 1\end{bmatrix} \right\}$.  
    \item The transition probability from a state $s=(n,\ell)$ to another state $s'=(n',\ell')$ is obtained for action $\bm{u}$ as
    \begin{align}
        T(s,s',\bm{u})=\begin{cases}
               \left(\bm{P}^\ell\right)_{nn'} & u(n')=\ell', \\
               0, & o \backslash w.
            \end{cases} \label{eq:Tp}
    \end{align}
    For instance, consider the state $s=(1,2)$, and the action $\bm{u}=\begin{bmatrix} 1 & 2 & 2 \end{bmatrix}$ for a process 
    \begin{align}
    \bm{P}=\begin{bmatrix}
    0.70 & 0.25 & 0.05 \\
    0.05 & 0.90 & 0.05 \\
    0.10 & 0.30 & 0.60        \end{bmatrix} .   
    \end{align}
    The probability of the next symbol is obtained from the first row of $\bm{P}^2$, which is 
    \begin{align}
       \big(\bm{P}^2\big)_{1*}=\begin{bmatrix}
               0.5075 &   0.4150  &  0.0775
        \end{bmatrix}.
    \end{align} 
    From \eqref{eq:Tp}, the transition probabilities for the next symbols can be calculated as
    \begin{align}
            T(&s=(1,2),s',\bm{u}=\begin{bmatrix}
            1 & 2 & 2
            \end{bmatrix})  \nonumber \\&=\begin{cases}
               0.5075, & s'=(1,1), \\
               0.4150, & s'=(2,2),\\
               0.0775, & s'=(3,2), \\
               0, & \text{otherwise}.
           \end{cases} \label{eq:Tp_ex}
    \end{align} 
    \item We finalize the MDP formulation by defining the cost function corresponding to the average transmission duration following the state $s=(n,\ell)$ and the action $\bm{u}$, 
    \begin{align}
        c(s=(n,\ell),\bm{u})&=\sum_{n'=1}^N T(s=(n,\ell),s=(n',u_{n'}),\bm{u})u_{n'} \nonumber \\ 
        &= \sum_{n'=1}^N\left(\bm{P}^\ell\right)_{nn'}u_{n'}. \label{eq:cost}
    \end{align}
\end{itemize}

A source coding policy $\phi:\mathcal{S}\to\mathcal{U}_N$ maps each state $s\in\mathcal{S}$ to an action $\bm{u}_s\in\mathcal{U}_N$, i.e., $\bm{u}_s=\phi(s)$. For a given policy $\phi$, the expected transmission time in the infinite horizon is
\begin{align}
    J^{\phi}=\sum_{s\in\mathcal{S}}\pi^{\phi}(s)c(s,\phi(s)), \label{eq:Jphi}
\end{align}
where $\pi^\phi(s)$ is the steady-state distribution of the state $s$ for given policy $\phi$. It satisfies the conditions $\pi^\phi(s)\bm{T}^{\phi}=\pi^\phi(s)$, and $\pi^\phi(s)\bm{1}=1$ for $|\mathcal{S}|\times|\mathcal{S}|$ transition matrix between states, $\left(\bm{T}^{\phi}\right)_{ss'}=T(s,s',\phi(s))$, and calculated by
\begin{align}
    \bm{\pi }^\phi & =\bm{1}^{\intercal}\left(\bm{T}^\phi+    \bm{1}\bm{1}^{\intercal}-\bm{I}\right)^{-1}. \label{eq:oneonetranspose}
\end{align}

We apply the policy iteration algorithm from \cite{ibe2013markov} to solve the defined MDP. The optimality equation for $s \in \mathcal{S}$ is
\begin{align}
    V_s&=\min_{  \bm{u}_s \in \mathcal{U}_N } V_s( \bm{u}_s)
    \\&=\min_{ \bm{u}_s \in \mathcal{U}_N}\left\{c(s,\bm{u}_s)-\eta +  \sum_{s' \in \mathcal{S} } T(s,s',\bm{u}_s) V_{s'} \right\}, \label{eq:opt_gen}
\end{align}
where $V_s$ denotes the optimal average cost starting at state $s$, $V_s(\bm{u}_s)$ is the average cost attained when action $\bm{u}_s \in \mathcal{U}_N$ is applied at initial state $s$ and then the optimal policy is applied, and $\eta=J^{\phi}$ denotes the long-term average cost. Notice that with this formulation, we have $|\mathcal{S}|$ equations with $|\mathcal{S}|+1$ unknown variables, including $\eta$. To solve these, we fix the value of a single state $s_0$, and obtain relative values of the remaining unknowns. The complete procedure is presented in Algorithm~\ref{alg:cap_ij}.

\begin{algorithm}
    \caption{Policy iteration algorithm }\label{alg:cap_ij}
    \begin{algorithmic}
        \State \textbf{Initialize:} Initiate all actions  $\bm{u}_s=\phi(s)$ $s \in \mathcal{S}$ with an arbitrary policy $\phi$.
        \State \textbf{Step 1: (MDP model)}  Obtain the values $T(s,s',\bm{u}_s)$ for each $s$ and $s'$. 
        \State \textbf{Step 2: (value determination)}: Obtain the long-term average cost $\eta$ and the relative values $V_s$ for $s\in\mathcal{S}\backslash s_0$ by fixing $V_{s_0}=0$ for $s_0=(1,1)$. Then solve the following $|\mathcal{S}|$ optimality equations 
        \begin{align}
            V_s =&c(s,\bm{u}_s) -\eta + \sum_{s'\in\mathcal{S}} T(s,s',\bm{u}_s) V_{s'},\quad s \in \mathcal{S}.
        \end{align}
        \State \textbf{Step 3: (policy improvement)}: For each $s\in\mathcal{S}$, obtain new policy by setting $\bm{u}_s$ as 
        \begin{align}
               \bm{u}_s=\underset{ \bm{u} \in \mathcal{U}_N}{\arg\min} \quad c(s,\bm{u})+\sum_{s'\in\mathcal{S}}T(s,s',\bm{u})V_{s'} \label{eq:pol}
        \end{align}
        \State \textbf{Step 4: (stopping rule)} If $|\eta^{(d)}-\eta^{(d-1)}|\leq \epsilon_\eta$ or $d=d_{\max}$ then stop. Otherwise, go to Step 1.  Here, $\eta^{(d)}$ denotes the long-time average cost obtained at iteration $d$.
    \end{algorithmic}
\end{algorithm}

\subsection{Extension to Continuous-Time Markov Chains}
In this part, we extend the proposed method for a continuous-time Markov chain (CTMC), $X(t)$, under the assumption that transmission of each bit takes $d$ sec. A CTMC is defined through its generator matrix $\bm{Q}$ with non-negative off-diagonal elements corresponding to the transition rates as,
\begin{align}
    \bm{Q}_{nn'}=\lim_{\varepsilon \to 0} \frac{Pr\left(X(t+\varepsilon)=n' \ | \ X(t)=n\right)}{\varepsilon},
\end{align}
and diagonal elements $\bm{Q}_{nn'}=-\sigma_n$ where $\sigma_n= \sum_{n'\neq n}\bm{Q}_{nn}$, making the row sums of $\bm{Q}$ zero. Whenever the process visits the state $n$, it stays there for an exponentially distributed time with parameter $\sigma_n$, then, a state transition to another state $n' \neq n$ occurs with probability $\rho_{nn'}=\frac{\bm{Q}_{nn'}}{\sigma_n}$ \cite{durrett1999essentials}. 

From Kolmogorov's forward equation \cite{durrett1999essentials}, we can obtain the state transition probabilities in unit time $d$ as
\begin{align}
    Pr(X(t+d)=n'|X(t)=n)=\Big(\mathrm{Exp}({\bm{Q}d})\Big)_{nn'}.
\end{align}
Then, we can obtain the optimal source coding policy for CTMC by applying the same procedure to the matrix $\bm{P}=\mathrm{Exp}({\bm{Q}d})$. 

\section{Numerical Results}
We compare the proposed policy with two benchmark policies. Both benchmark policies are based on the well-known Huffman coding \cite{huffman1952method}, which is known to be optimal in the sense of minimizing the expected transmission duration of the next symbol with known occurrence probabilities, and all Huffman codes are complete codes \cite{cover1999elements}. We denote a function $h(\bm{p})=[h_1(\bm{p}),h_2(\bm{p}),\dots,h_N(\bm{p})]$ that generates the codeword length of the next symbols by applying the Huffman code to the occurrence probabilities, which is denoted by $\bm{p}$.

\subsection{Benchmark Policies}
\paragraph{Myopic Huffman Policy} In this policy, for each state $s=(n,\ell)$ we apply Huffman coding for the next symbol $n'\in\mathcal{N}$ by using probabilities $(\bm{P}^\ell)_{nn'}$. This method only minimizes the expected transmission length for the next symbol; hence, it is a myopic policy. The policy $\phi^{m}$ for this benchmark can be expressed as
\begin{align}
    \phi^{m}(n,\ell)=h\left(\big(\bm{P}^\ell\big)_{n*}\right),
\end{align}
where $\big(\bm{P}^\ell\big)_{n*}$ corresponds to the probability of the next symbol after state $s=(n,\ell)$.

\paragraph{Steady-State Huffman Policy} In this policy, we consider a static mapping between each state $n\in\mathcal{N}$ to a codeword with length $\ell_n$ based on the steady-state distribution of the symbol. Regardless of the last transmitted symbol and the duration of this transmission, the next symbol $n'$ is encoded with that codeword. Similar to \eqref{eq:oneonetranspose}, we first calculate the steady-state distribution of all symbols with
\begin{align}
    \bm{\pi}^{st}&=\bm{1}^T(\bm{P}+\bm{1}\bm{1}^T-\bm{I}).
\end{align}
Then, we obtain a codebook by applying Huffman coding to these probabilities, and the policy $\phi^{st}$ can be expressed as
\begin{align}
    \phi^{st}(n,\ell)=h(\bm{\pi}^{st}).
\end{align}

We denote the average transmission lengths obtained by our proposed policy and the benchmark methods with $L^*$, $L^m$, and $L^{st}$, respectively, which can be calculated by \eqref{eq:Jphi}.

For the optimal policy, we initiate the algorithm with \emph{myopic Huffman policy} as the initial policy. Parameters of the stopping rule are selected as $\epsilon=10^{-4}$ and $d_{\max}=30$. In the policy improvement step, we exhaustively search for a complete code that minimizes \eqref{eq:pol}. Table~\ref{tab:comp} gives the number of complete codes for the alphabet sizes between $3\leq N \leq 8$.

\begin{table}[h]
    \centering    
    \caption{The number of possible complete codes, equivalently the action space size $|\mathcal{U}_N|$, for $3\leq N \leq 8$.}
    \begin{tabular}{c @{\hspace{1em}}  c c c c c c}
        \toprule
        $N$ & 3 & 4 & 5 & 6 & 7 & 8 \\ 
        \midrule
        $|\mathcal{U}_N|$ & 3 & 13 & 75 & 525 & 4347 & 41245 \\
        \bottomrule
    \end{tabular}
    \label{tab:comp}
\end{table}

\subsection{Generation of Remote Markov Sources} 
In our simulations, we consider the linear combinations of the following transition Markov processes.

\begin{enumerate}[label=\roman*)]
    \item \emph{Markov Chains with Homogeneous Transition Matrices:} In this case, self-transition probabilities for each symbol $n$ are the same and equal to $\alpha\in(0,1)$, and the transition probability between symbols $n$ and $n'\neq n$ are $\frac{1-\alpha}{N-1}$. Transition matrix for this process is denoted by $\bm{H}^{(\alpha)}$, and it can be expressed in the matrix notation as 
    \begin{align}
        \bm{H}^{(\alpha)}=\left(\alpha-\dfrac{1-\alpha}{N-1}\right)\bm{I}+\left(\dfrac{1-\alpha}{N-1}\right)\bm{1}\bm{1}^{\intercal}. \label{eq:matH}
    \end{align}
    \item \emph{Markov Chains with Randomly Chosen Transition Matrices:} In this case, the rows of the transition matrices are generated independently and randomly. We denote the transition matrix for this process by $\bm{R}$, and the elements of this matrix are generated in the following way. First, each element of the matrix is generated from the uniform distribution with $\tilde{\bm{R}}_{nm}\sim U(0,1)$, $n,m\in\mathcal{N}$. Then, the matrix $\tilde{\bm{R}}$ is normalized as
    \begin{align}
        {\bm{R}}_{nm}=\dfrac{\left(\tilde{\bm{R}}\right)_{nm}}{\sum_{m'\in\mathcal{N}} \tilde{\bm{R}}_{nm'}}. \label{eq:matR}
    \end{align}
\end{enumerate}

\begin{figure}
    \centering
    \includegraphics[width=0.75\linewidth]{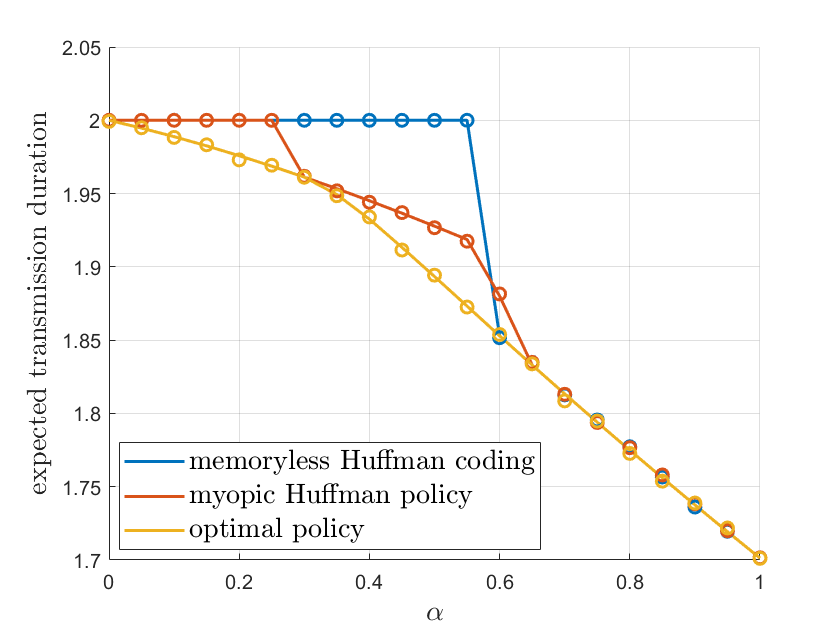}
    \caption{Analytical and simulation results for a random matrix realization $\bm{R}_0$ in \eqref{eq:R0}, with transition probability matrix $\bm{P}$ calculated according to \eqref{Pcalc}.}
    \label{fig:simN4}
\end{figure}

\subsection{Experiments and Simulation Results}
In the first experiment, we consider a Markov chain with $N=4$ states, with a transition matrix given as 
\begin{align}
    \bm{P}=(1-\beta) \bm{H}^{(0.5)}+\beta \bm{R}_0, \label{Pcalc}
\end{align}
where $\bm{H}^{(0.5)}$ is obtained from \eqref{eq:matH} for $\alpha=0.5$, and $\bm{R}_0$ is one of the realizations of \eqref{eq:matR}, which is
\begin{align}
    \bm{R}_0=\begin{bmatrix}
    0.1426 & 0.4996 & 0.0409 & 0.3169 \\
    0.3542 & 0.5398 & 0.0858 & 0.0202 \\
    0.1732 & 0.3522 & 0.0946 & 0.3800 \\
    0.1124 & 0.3401 & 0.2936 & 0.2540
    \end{bmatrix}. \label{eq:R0}
\end{align}

Fig.~\ref{fig:simN4} illustrates the results of this setting for varying $\beta$ values. We obtain analytical results from \eqref{eq:Jphi} and \eqref{eq:oneonetranspose}, and illustrate them with solid lines. For simulation results, which are illustrated with circles, we initiate the Markov chain from state $1$, and calculate the average transmission duration over $10^6$ transmissions. The agreement between simulation and analytical results verifies our analysis. For this realization, we observe that the performances of the policies are in the order of $L^{st}\geq L^{m} \geq L^*$ except for $\beta=0.6$ when the steady-state Huffman policy outperforms the myopic-Huffman policy. In addition, the average transmission durations for all policies become equal to each other when $\beta=0$ (the transition matrix is purely homogeneous with $\alpha=0.5$) and after $\beta\geq0.65$ (the transition matrix is close to $\bm{R}_0$). 

\begin{figure}
    \begin{center}
    \subfigure[]{\includegraphics[width=0.45\linewidth]{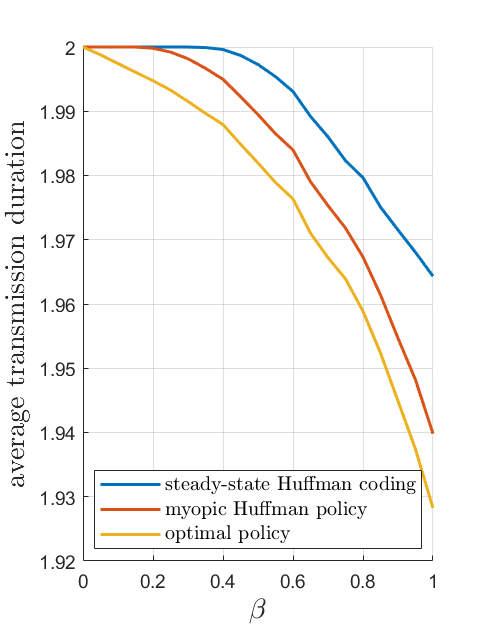}} ~ 
    \subfigure[]{\includegraphics[width=0.45\linewidth]{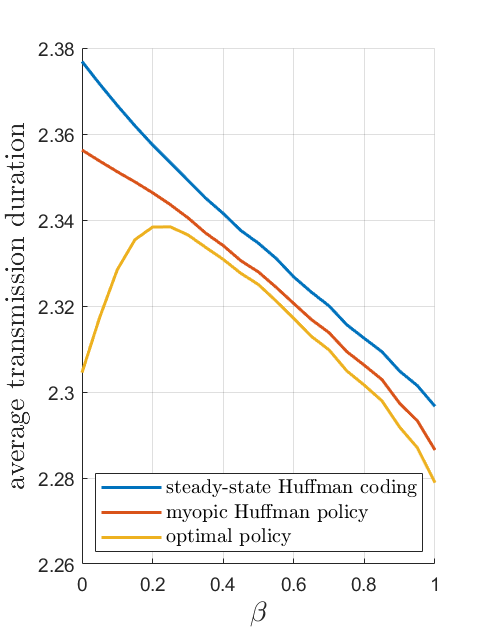}} 
    \end{center}  
    \caption{Comparison of the average transmission durations for state transition matrices generated randomly according to \eqref{eq:second_exp} for alphabet sizes, i.e., number of states in the Markov chain, (a) $N=4$ and (b) $N=5$.}
    \label{fig:example2}
\end{figure}

In the second experiment, illustrated in Fig.~\ref{fig:example2}, we repeat the previous experiment for $10^4$ state transition matrices generated according to
\begin{align}
    \bm{P}=(1-\beta) \bm{H}^{(0.5)}+\beta \bm{R}, \label{eq:second_exp}
\end{align}
and average the transmission durations over all processes. We observe that the order between policies is  $L^{st}\geq L^{m} \geq L^*$ consistently for all $\beta$. Another result of this experiment is that, we reach a lower average transmission duration even for the homogeneous process $\bm{H}^{(0.5)}$ for $N=5$; see Fig.~\ref{fig:example2}(b), $\beta=0$.

\begin{figure}
    \begin{center}
    \subfigure[]{\includegraphics[width=0.45\linewidth]{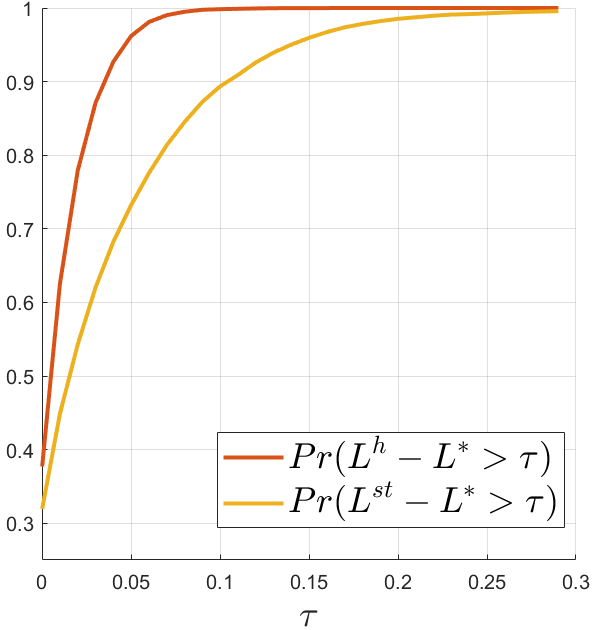}} ~ 
    \subfigure[]{\includegraphics[width=0.45\linewidth]{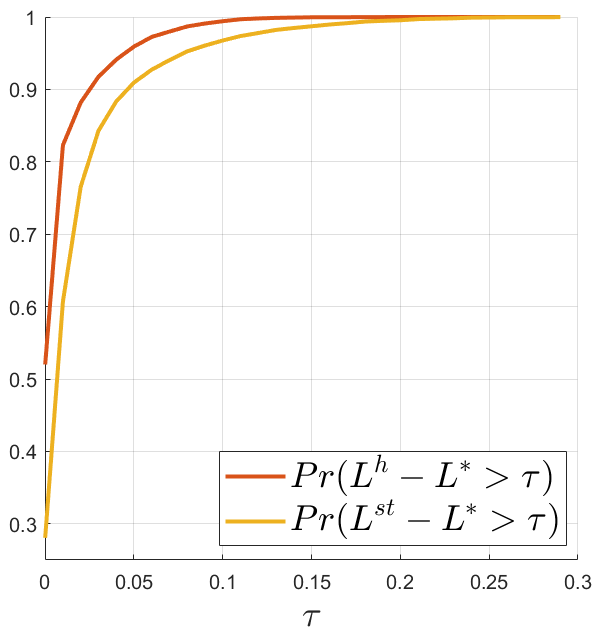}}  
    \end{center}  
    \caption{The probability that the average transmission duration difference between the proposed method and the benchmark methods is less than or equal to a threshold $\tau$ for randomly generated state transition matrices for Markov chains with number of states equal to (a) $N=4$ and (b) $N=5$.}
    \label{fig:outage}
\end{figure}

In the next experiment, illustrated in Fig.~\ref{fig:outage}, we investigate the performance gain of the optimal policy compared to the benchmark policies for $N=4$ and $N=5$. For each case, we generate policies for $10^4$ random realizations of the state transition matrices, which are obtained by \eqref{eq:matR}. Then, we calculate the probability that the performance gain of using the optimal policy is larger than a varying threshold $\tau$ compared to the benchmark policies by calculating over all processes, denoted as $Pr(L^{k}-L^*>\tau)$, $k=\{m,st\}$. The experimental results suggest that for randomly generated transition matrices, the benchmark policies can have near-optimal performance, that is, the performance gain of the optimal policy can be relatively small. For instance, we observe that the performance of our proposed algorithm is close to the performance of the myopic Huffman policy with probability $\sim 0.52$, and $\sim 0.48$, for $N=4$ and $N=5$, respectively. However, they also indicate that the performance gain may be significant with a non-negligible probability.

Finally, Table~\ref{tab:N} illustrates the expected values of the average transmission durations and the performance gain of the optimal policy compared to benchmark polices. In this experiment, we again generate the state transition matrices randomly according to \eqref{eq:matR} for different $N$ values. We observe that the optimal policy consistently reduces the expected value of the average duration regardless of the alphabet size. Additionally, we observe that the performance gain of the optimal policy monotonically decreases with alphabet size for steady-state Huffman policy, and it is the largest when $N=4$ for myopic Huffman policy. This table should be read together with the results from Fig.~\ref{fig:outage}, which indicates that benchmark policies can be near-optimal for some cases.

\begin{table}
    \centering
        \caption{Comparison of policies for Markov chains with randomly generated state transition matrices for different numbers of states $N$.}
    \begin{tabular}{ccccc}
        &  $N=3$ &  $N=4$ & $N=5$  &  $N=6$\\
        $\mathbb{E}[L^{st}]$ & $1.5629$  & $1.9613$ & $2.2956$ & $2.5810$ \\
        $\mathbb{E}[L^{m}]$ & $1.5152$ & $1.9378$ & $2.2860$ &  $2.5750$\\
        $\mathbb{E}[L^{*}]$ & $1.5122$  & $1.9252$  & $2.2779$ & $2.5740$ \\
        $\mathbb{E}[L^{st}-L^*]$ & $0.0506$ & $0.0361$ & $0.0177$ & $0.0069$ \\
        $\mathbb{E}[L^{m}-L^*]$ & $0.0030$ & $0.0126$ & $0.0081$ & $0.0009$ \\
    \end{tabular}
    \label{tab:N}
\end{table}

\section{Conclusion and Future Directions}
We studied the source coding problem for real-time remote monitoring of a Markov chain when the state transition time durations and bit transmission time durations are of the same scale. We formulated the problem as an MDP by using the augmented states approach, and proposed a policy iteration algorithm to obtain an optimal policy. In addition, we extended our work to CTMCs for integer-valued codeword lengths and constant transmission delays per bit. We additionally analyzed two Huffman code-based benchmark policies and compared their performance with our proposed optimal policy. Our results show that the proposed optimal policy always performs better than (or equal to) the benchmark policies. 

The scope of this paper has been limited to a smaller alphabet size (number of states in the Markov chain) and integer-valued codeword lengths. That ensures that the size of the action space for this problem is suitable for exhaustive search in the policy improvement step. Thus, the extension of our work for continuous-valued codeword lengths, or larger alphabet sizes, require further attention. Furthermore, recent works \cite{Yates__HowOftenShouldone, yates2020age, optimal_codes} discuss that the transmission duration may not be the best metric for real-time monitoring problems. Another direction of our work is to find the optimal policies minimizing a semantic or a freshness metric instead of the average transmission duration.

\bibliographystyle{IEEEtran}
\bibliography{bibl}
\end{document}